\pdfoutput=1
\PassOptionsToPackage{hyperref}{
	hidelinks,
	pdfusetitle,
	extension={},
	breaklinks=true
}
\documentclass[conference]{IEEEtran}

\newif\ifFinalCopy
\FinalCopytrue

\newif\ifIEEEXplore
\IEEEXplorefalse

\usepackage[T1]{fontenc}
\usepackage[utf8]{inputenc}
\usepackage{csquotes}
\usepackage{environ}
\usepackage{ifthen,xspace,xparse}

\usepackage{array,amsmath,amsfonts,amssymb}
\usepackage[
	style=ieee,
	isbn=false,
]{biblatex}
\usepackage{enumitem}
\usepackage{float,graphicx,tabularx,colortbl,booktabs,listings}
\usepackage[caption=false,font=footnotesize]{subfig} 
\usepackage[colorinlistoftodos, textsize=footnotesize,%
	\ifCLASSOPTIONpeerreview disable,%
	\else\ifFinalCopy disable,%
	\fi\fi
]{todonotes}
\usepackage{siunitx}
\usepackage{tikz,pgfplots,pgfplotstable,pgfpages}
\usetikzlibrary{positioning,shapes,math,calc,arrows.meta}
\usepackage[capitalize,nameinlink]{cleveref} 
\usepackage{balance} 

\tikzstyle{square}    = [regular polygon,regular polygon sides=4]
\tikzstyle{inuse}     = [black,text=gray!10]
\tikzstyle{unused}    = [gray!40]
\tikzstyle{PEBlock}   = [draw,square,inner sep=0.15]
\tikzstyle{DFGNode}   = [draw,black,circle,minimum width=.75,inner sep=.25]
\tikzstyle{DFGEdge}   = [semithick]
\tikzstyle{MRRGNode}  = [draw,black,circle]
\tikzstyle{MRRGFUNode} = [MRRGNode, very thick]
\tikzstyle{MRRGEdge}  = [semithick, rounded corners=1mm]
\tikzstyle{MRRGExtEdge} = [MRRGEdge,dotted]

\tikzstyle{skipconRight} = [-,looseness=0.4,bend right=45]
\tikzstyle{skipconLeft} = [-,looseness=0.4,bend left=45]

\DeclareMathOperator*{\compatnodes}{comp}
\DeclareMathOperator*{\neighbours}{neigh}
\DeclareMathOperator*{\pathsfor}{paths}
\DeclareMathOperator*{\graphoutputs}{outputs}

\DeclareMathOperator*{\compatpaths}{pcompat}

\def\edgevar{\mathbf e}

\newcommand{\questeq}{\stackrel{?}{=}}

\def\cgrame@{CGRA-ME}
\def\adres@{ADRES}
\def\clustered@{Clustered}
\def\hycube@{HyCUBE}
\def\fmax@{\(f_{\text{max}}\)}
\def\iint@{II}
\def\nei@{neighbour}
\def\pnei{neighbour} 
\def\spellBehaviour@{behaviour}
\def\spellCentre@{centre}
\def\numnei@{number-of-\nei@}
\def\Numnei@{Number-of-\nei@}
\def\relaxedPathConstraintPlacementName@{relaxed-constraint-placement}

\def\heurTimeNinteyP@{410}
\def\geomeanSpeedup@{37.6}
\def\geomeanSpeedupNoTimeout@{5.88}
\def\arithMeanSpeedup@{861}
\def\medianSpeedup@{41.7}

\renewcommand*{\bibfont}{\footnotesize} 

\def\lastreferencepage{9} 

\def\balanceissued{unbalanced}
\renewbibmacro{finentry}{
	\ifnum\thepage=\lastreferencepage%
		\expandafter\ifx\expandafter\relax\balanceissued\relax%
		\else%
			\balance%
			\gdef\balanceissued{\relax}%
		\fi%
	\fi%
	\finentry%
}

\pgfplotsset{
	compat=1.14, 
	table/col sep=tab,
	scale only axis,
	small,
	width=0.7\linewidth,
}

\pgfplotstableset{my formatting/.style={
	every head row/.style={
		before row=\toprule,
		after row=\midrule,
	},
	every last row/.style={after row=\bottomrule},
	empty cells with={---},
}}

\newcommand\textvtt[1]{{\normalfont\fontfamily{cmvtt}\selectfont #1}} 

\newcounter{constraintCounter}
\DeclareRobustCommand{\ILPConstraint}[1]{\refstepcounter{constraintCounter}\emph{\theconstraintCounter) #1}}
\crefname{constraintCounter}{constraint}{constraints}

\title{Generic Connectivity-Based CGRA Mapping via Integer Linear Programming}

\author{
	\ifCLASSOPTIONpeerreview (Authorship Removed for Blind Review)\else 
	\texorpdfstring{
		\IEEEauthorblockN{Matthew J. P. Walker}
		\IEEEauthorblockA{%
			Edward S. Rogers Sr. Department of
			Electrical and\\ Computer Engineering,
			University of Toronto\\
			Toronto, Ontario, Canada\\
			Email: matthewjp.walker@mail.utoronto.ca%
		}
		\and
		\IEEEauthorblockN{Jason H. Anderson}
		\IEEEauthorblockA{%
			Edward S. Rogers Sr. Department of
			Electrical and\\ Computer Engineering,
			University of Toronto\\
			Toronto, Ontario, Canada\\
			Email: janders@ece.utoronto.ca%
		}
	}{
		Matthew J. P. Walker <matthewjp.walker@mail.utoronto> and Jason H. Anderson <janders@ece.utoronto.ca>
	}%
	\fi
}

\begin{document}
\ifIEEEXplore
	\begin{NoHyper}
\fi

\maketitle

\begin{abstract}
Coarse-grained reconfigurable architectures (CGRAs) are
	programmable logic devices with large coarse-grained ALU-like logic blocks,
	and multi-bit datapath-style routing.
CGRAs often have relatively restricted data routing networks,
	so they attract CAD mapping tools that use exact methods,
	such as Integer Linear Programming (ILP).
However,
	tools that target general architectures
	must use large constraint systems
	to fully describe an architecture's flexibility, resulting in lengthy run-times.
In this paper,
	we propose to derive connectivity information from an otherwise generic device model,
	and use this to create simpler ILPs,
	which we combine in an iterative schedule
	and retain most of the exactness of a fully-generic ILP approach.
This new approach
	has a speed-up geometric mean of \geomeanSpeedupNoTimeout@\(\times\) when considering benchmarks that
	do not hit a time-limit of 7.5 hours on the fully-generic ILP,
	and \geomeanSpeedup@\(\times\) otherwise.
This was measured using the set of benchmarks used to originally evaluate the fully-generic approach
	and several more benchmarks representing computation tasks, over three different CGRA architectures.
All run-times of the new approach are less than 20 minutes, with 90th percentile time of \heurTimeNinteyP@ seconds.
The proposed mapping techniques are integrated into,
	and evaluated using the open-source CGRA-ME architecture modelling and exploration framework~\cite{CGRAME-intro}.
\end{abstract}

\ifCLASSOPTIONpeerreview
	\begin{center} \Large For Peer Review \end{center}
\section*{Change Log}
\subsection{2019-03-24}
\subsubsection{Changes Requested}
The reviewers note that this paper needs major revision to properly
	present the approach and results.  In particular, there is need for more
	deliberate (and where appropriate formal) description of the challenges and
	solutions including more careful and precise statement of the definitions
	and more careful (less informal) description of key components.
\subsubsection{Major Changes Made}
\begin{itemize}
	\item \cref{sec:the-idea} has been heavily edited to be more precise and give more detail of the approach.
	\item Some clarification has been added to the opening text of \cref{sec:ilp-details} that describes the ILP variables.
	\item \cref{fig:nn-mapfrac} replaces a table, and gives more information about mappability at different \numnei@
\end{itemize}
\subsubsection{Comments}
Although the exact changes requested were not clear to us,
	we hope these changes make the paper acceptable,
	and are willing further to iterate with the Shepherd.
\subsection{2019-03-27}
\subsubsection{Changes Requested}
It is not clear how to interpret Fig. 1b.
	When the a node (add here) has two outputs, are they two copies of the same output, or
	are they different outputs?
    Are there limits on the loops, or do they run forever?
    How are state values (accumulations) initialized?

"A typical implementation of a PE will contain one FU node for computation
	connected to registers, and input and output crossbars, such as Fig. 2c.”
	But Fig. 2c shows two FUs. Is that really one FU at different times? Also
	where are the input and output crossbars?  Caption talks about muxes and
	the nodes are labeled with "m".

"except that paths starting at the same MRRG vertex may initially overlap."
	--> say why.  Is this a statement about your problem being one of
	hypergraphs (allowing fanout) while the original is about graphs (no
	fanout).  State that (or whatever it is) more directly.

"then at least one of it’s fanout must be too.”
	Maybe show the equational constraint that goes with this to be precise.

"if two nodes are connected by an edge at most one may have multiple
	fanins" --> It's not clear what this is about.
   Maybe useful to show an example of what can go wrong here?
   Maybe useful to show the equational constraint that goes with this.

IIIA -- make connection of this "fanin of the operation must be mapped to
	neighboring PE" to the earlier statement/setup in intro about destinations
	being a limited number of hops away.

IIIB "from the driver to the fanout" --> is "fanout" used here to simply
	refer to the "sink" of one FU driving another.  Or is it being used to
	refer to there being multiple sinks on a multi-point net?
	If it's about multiple sinks, something seems to be missing.
	If it's about a destination of a 2-point net, maybe use a different term
	since you use the term "fanout" elsewhere to be talking about multiple
	sinks.

End of IIIB -- could use an example (probably a figure) to ground what the final paragraph is describing.

IIIC "In this work we use the constraint-based approach because it overall takes less time to discover a placement that will route.”
	Which one of the two alternatives described is the “constraint-based approach?” Neither one use the term constraint to help guide the reader.

I am confused as to which is constraint based and which is cost based and which one was chosen.

Fig. 8 -- why do the curves have different numbers that occur to the same
   percentages?   Do the different architectures have different number of
   benchmarks>?

VA "different solver seeds"  --> seeds imply randomness?
     What aspect of the solution is driven by random selection?

\subsubsection{Changes Made}
\begin{itemize}
	\item
		``re-convergence'',
		context IDs in figures, ``it's'', ``and ILP'', ``Or,'',
		``in in'', ``map ... map'', ``are ... are'', ``arch/II'', `[4]To extend'.
	\item
		Mention constraint numbers for principles,
		DFGs are hypergraphs,
		MRRG is flattened,
	\item
		Add more detail about existing approach,
		clarify 'constraint-based approach',
		clarify Fig. 8 caption,
		mention seed affects runtime,
\end{itemize}
\subsubsection{Comments}
Thank you for taking a close look at our paper,
	and providing detailed feedback.
We believe that we have addressed all points,
	except for the figure for Section III-B,
	due to lack of space.

\subsection{2019-03-29}
\subsubsection{Changes Requested}
Constraint:
	A: First paragraph of III.C describes one scheme (combined placement and routing)
	B: Second paragraph of III.C describes another scheme (split placement and  routing)
	C: Third paragraph of III.C describes another scheme (ILP cost function for placement)
		(and notice that it talks about "adding the constraints for  choosing paths",
		while the word "constraint" was never used in either of the
		previous paragraphs)

	Later in the 3rd paragraph, "constraint-based approach" is used for the
	first term.  Which of the 3 cases is this?

	Later in 3rd paragraph, "cost-based approach" is used.  I'm guessing
	that's C since it has a "cost function", but then again C is the only
	one that uses the "constraint" term in its description.

	In Section IV before "1)" it talks about:
		placement-only ILP
		routing-only ILP

	Section V.B, paragraph 3 mentions the "placements-with-congestion ILP".

	I'll go out on a limb and guess:
		cost-based is C
		constraint-based approach is B
		placement-only ILP is part of B
		routing-only ILP is part of B
		placement-with-congestion ILP is C

	The reader shouldn't be required to decipher which of these various
	terms you are using are actually equivalent.

	If my guess is correct, start by introducing the "constraint-based" term in
	paragraph B.  For example, ther last sentence could become:
		"This approach, which we call constraint-based, has small ILPs..."
	OR
		"This constraint-based approach has small ILPs..."

	If placement-with-congestion ILP is a better description, maybe use it
	instead of "cost-based" and introduce it in the 3rd paragraph.
		"To address this, one solution is a placement-with-congestion scheme
		where we choose an ILP cost function..."
	then
		"This is true even for the placement-with-congest scheme..."

Seed:
   First (now only?) mention of seed is Section V.A, first paragraph.
   It remains unclear what randomness is involved that uses a seed.

\subsubsection{Changes Made}
\begin{itemize}
	\item ``Conversely'', ``may be have'', ``graph. [4].'', ``an neighbouring'', ``data that is''.
	\item seed: clarify it is the ILP solver's seed.
	\item Name approaches, and use the names more consistently. Eg. ``relaxed-constraint-placement'' replaces ``constraint-based''.
\end{itemize}
\subsubsection{Comments}
	In the previous version, ``constraint-based'' and ``cost-based'' referred to different choices for the placement ILP.
	We hope the new wording is more understandable.

\fi

\IEEEpeerreviewmaketitle

\section{Introduction}
Coarse-grained reconfigurable architectures (CGRAs) are
	a class of programmable logic device where the processing elements (PEs) are
	large ALU-like logic blocks, and the interconnect fabric is bus-based.
This stands in contrast to field-programmable gate arrays (FPGAs),
	which are configurable at the individual logic-signal level.
CGRAs dedicate less area to flexibility/programmability,
	and require far fewer configuration bits than FPGAs,
	thereby easing CAD complexity by reducing the number of decisions tools need to make.
Despite their reduced flexibility,
	CGRAs are an ideal media for applications where:
	1) \emph{some} flexibility is required,
	2) software programmability is desired,
	and
	3) compute/communication needs closely match with the CGRA capabilities.
CGRAs can be realized as custom ASICs,
	or alternatively, implemented on FPGAs as overlays,
	and a number of commercial
	and academic
	architectures have been proposed, stretching back to the 1990s~\cite{DecadeOfReconfComp,ReconfCompArchsTessier}.
With the coming end to Moore's Law,
	CGRAs are receiving renewed interest as platforms for domain-specific compute acceleration.
	\todo{cite?}
As such,
	it is desirable to develop methodologies for
	the modelling and evaluation of hypothetical CGRAs.
The open-source CGRA-ME (CGRA Modelling and Exploration) framework from
	the University of Toronto~\cite{CGRAME-intro} aims to provide this capability.

The CGRA-ME framework allows a human architect to describe
	a hypothetical CGRA using an expressive graph-based device model,
	and provides generic mapping approaches that allow
	an application benchmark to be mapped into the described CGRA.
Of particular interest for architecture exploration is
	the integer linear programming-based (ILP) mapping approach,
	as it provides certainty regarding the mappability of an application benchmark into an architecture
	\cite{CGRAME-ILP}.
With such an exact mapper,
	an architect can be confident whether an architecture is viable
	for a set of applications.
This approach performs well for very small architectures and application benchmarks,
	but does not scale very well
	-- the runtimes of larger benchmarks and architectures can
	extend into the day range on a typical workstation.
To truly enable architecture exploration,
	mapping times should be significantly, and consistently, less.
It is precisely this challenge we address in this paper, namely,
	that of providing scalable mapping algorithms for CGRAs,
	while retaining the exactness property and genericity.

Through use of \cgrame@, we have observed that
	mapping times are seemingly random -- mapping slightly perturbed benchmarks
	to the same architecture may take seconds, minutes or hours.
This effect becomes more pronounced with larger ILP models,
	the size of which is determined by both benchmark and device model size.
Application benchmarks are generally quite small,
	and cannot be simplified.
Conversely, the device model generally has thousands of vertices,
	leading to a large ILP problem.

The existing technique \cite{CGRAME-ILP} retains all the flexibility of the device model,
	however, some of this flexibility is effectively useless.
For example,
	many CGRAs are grid-based,
	and we observe that is is uncommon for the output of one PE to
	have a destination PE that is more than two ``hops'' away.
Also,
	many CGRAs have extremely restricted routing networks,
	where, for example, only nearest-\nei@ connectivity is present between PEs.

Given limited practical need for long routing paths,
	and generally non-contested connections,
	we present a simpler, smaller, ILP that captures most of the flexibility (\cref{sec:ilp-details}).
Further,
	PEs that are ``close together'',
	and the connections between them,
	can be derived from \cgrame@'s device model
	(\cref{sec:connectivity-intro,sec:paths-intro}) and the ILP formulation can be restricted to consider such information.
Finally,
	\cref{sec:composition-intro} presents an algorithm to efficiently map benchmarks to CGRAs by
	using variants of the proposed ILP formulation, where we iteratively generate ILP mapping formulations that consider successively larger portions of the solution space.
	\todo{word this sentence a better}

\section{Background}

\subsection{Data-Flow Graphs}
A benchmark or application kernel's
	essential structure can be represented as a directed hypergraph,
	called a \emph{data-flow graph} (DFG),
	such as those in
	\cref{fig:sample-dfgs}.
In simple cases (e.g. \cref{fig:simple-dfg}) they may be thought of as similar to abstract syntax trees,
	but only include values (corresponding to edges) and
	operations (corresponding to vertices).
In more more complex cases they may have loops (e.g. \cref{fig:complex-dfg})
	and re-convergent paths (e.g. \((a+b)*(a+c)\)).
For the purposes of CGRA mapping, the edges are interpreted as dependency relations between operations,
	capturing the set of operations that must be performed before
	a given operation can proceed, and where data must be routed.
Loads, stores, inputs, outputs and constants are also modelled as vertices,
	and loop-carried dependencies correspond to back-edges/loops.
The input to mapping CGRA-ME is: 1)
	an application DFG,
	and 2)
	a device model for the targeted CGRA,
	called a Modulo Routing Resource Graph (MRRG),
	described in \cref{sec:mrrgs}.

\begin{figure}
	\centering
	\small
	\subfloat[\textvtt{a = b * (c + d)}]{\label{fig:simple-dfg}
		\begin{tikzpicture}
			\node[draw, rectangle]                             (out)  {output a};
			\node[draw, circle,    above      =.25 cm of out]  (mul)  {mul}       edge[->,thick] (out);
			\node[draw, circle,    above right=.25 cm of mul]  (add)  {add}       edge[->,thick] (mul);
			\node[draw, rectangle, above left =.25 cm of mul]  (in3)  {input b}   edge[->,thick] (mul);
			\node[draw, rectangle, above      =.75 cm of add]  (in1)  {input d}   edge[->,thick] (add);
			\node[draw, rectangle, above left =.25 cm of add]  (in2)  {input c}   edge[->,thick] (add);
		\end{tikzpicture}
	}
	\hspace{.25cm}
	\subfloat[A sum over an array of 4-byte numbers]{\label{fig:complex-dfg}
		\begin{tikzpicture}
			\node[draw, rectangle]                             (out)  {output sum};
			\node[draw, circle,    left       =.25 cm of out]  (add)  {add}         edge[->,thick,loop left] () edge[->,thick] (out);
			\node[draw, circle,    above      =.25 cm of add]  (ld)   {load}        edge[->,thick] (add);
			\node[draw, circle,    above      =.25 cm of ld]   (off)  {add}         edge[->,thick] (ld);
			\node[draw, rectangle, above left =.25 cm of off]  (adr)  {input addr}  edge[->,thick] (off);
			\node[draw, circle,          right=.25 cm of off]  (cnt)  {add}         edge[->,thick,loop above] () edge[->,thick] (off);
			\node[draw, circle,    below      =.25 cm of cnt]  (inc)  {4}           edge[->,thick] (cnt);
		\end{tikzpicture}
	}
	\caption{Example DFGs.}\label{fig:sample-dfgs}
\end{figure}
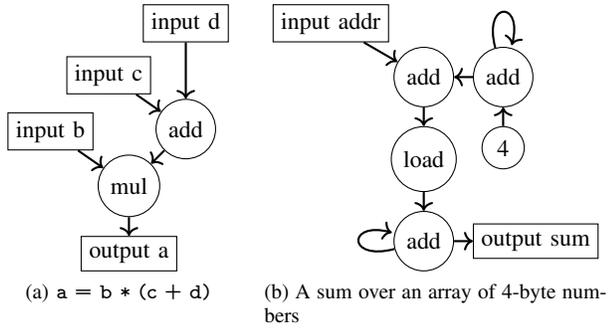

\subsection{Multi-Context CGRAs}

An important property of previously proposed CGRAs (e.g.~\cite{DRESC,ADRES-long,SYSCORE,HyCUBE-intro}) is
	the notion of \emph{multiple contexts}.
A context is a single configuration of the CGRA's logic functionality and routing connectivity.
A two-context CGRA would contain \emph{two} copies of its configuration cells:
	configuration 0 and 1.
The typical \spellBehaviour@ of such a CGRA is
	to cycle between the two contexts on a cycle-by-cycle basis.
This implies that the hardware functionality and routing can change each cycle.
The hardware is thus ``time multiplexed'',
	where PEs and routing can be used for different purposes in each context.
A PE can, for example, perform an addition in context 0,
	store the result at the clock edge, and then perform a multiply in context 1.
This is as opposed to today's commercial FPGAs, which are single context.
A typical CGRA has a range of configuration context counts
	that it can physically realize,
and a mapping tool will typically try to minimize this,
	as it is equal to the initiation interval (\iint@) -- the rate at which new inputs are consumed by the CGRA.

\subsection{Modulo Routing Resource Graphs (MRRGs)}\label{sec:mrrgs}
An MRRG \cite{DRESC} is a graph data structure that is commonly used to
	model the hardware connectivity and capability of CGRAs \cite{EcMS,GraphMinorCGRA}.
It is used as the CGRA device model within the CGRA-ME framework \cite{CGRAME-ILP}.
An MRRG is a directed graph where a vertex is a hardware element in time and space,
	and an edge represents a possible fanout.
In the variant used by \cgrame@, a vertex is a 2-tuple \((s,t\)) with
	physical node id \(s\),
	and context number \(t\) (\textbf{t}ime).
Edges represent connectivity across time and space,
	and include connections that ``wrap around'' from \(t\) to \(t' \leq t\)
	(eg. \cref{fig:reg-mrrg}).
This is sufficient to describe the structure,
	but to capture the entire \spellBehaviour@ some extra data is tagged on each vertex,
	such as latency and its supported computation operations, if any.
We will refer to nodes that support computation operations as functional unit (FU) nodes.

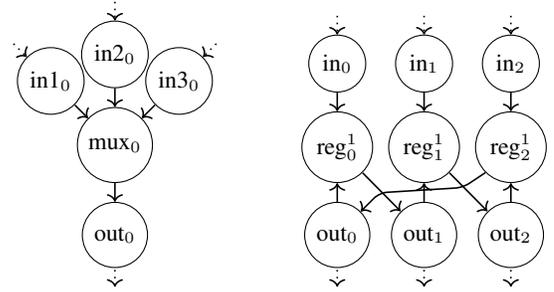
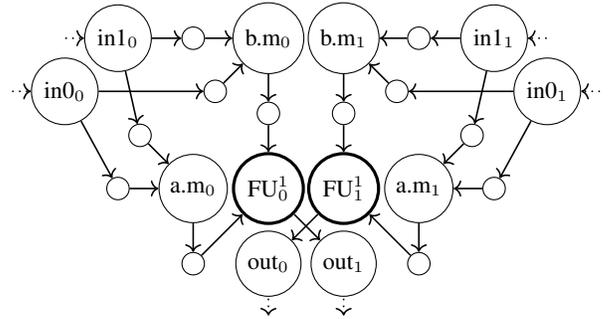
\begin{figure}\centering
	\small
	\begin{minipage}{0.9\linewidth}
	\subfloat[%
		A 3-to-1 mux for \iint@ = 1. The structure would be duplicated for higher \iint@,
		with appropriate changes to the \(t\) values.
	]{
		\label{fig:mux-mrrg}
		{

\begin{tikzpicture}
	\node[MRRGNode,                              ] (out) {\(\text{out}_0\)};
	\node[MRRGNode,     above      =.25 cm of out] (mux) {\(\text{mux}_0\)} edge[->,MRRGEdge       ] (out);
	\node[MRRGNode,     above left =.25 cm of mux] (in1) {\(\text{in1}_0\)} edge[->,MRRGEdge       ] (mux);
	\node[MRRGNode,     above      =.25 cm of mux] (in2) {\(\text{in2}_0\)} edge[->,MRRGEdge       ] (mux);
	\node[MRRGNode,     above right=.25 cm of mux] (in3) {\(\text{in3}_0\)} edge[->,MRRGEdge    ] (mux);
	\node[              below      =.25 cm of out] (oar) {}                 edge[<-,MRRGExtEdge] (out);
	\node[              above left =.25 cm of in1] (an1) {}                 edge[->,MRRGExtEdge] (in1);
	\node[              above      =.25 cm of in2] (an2) {}                 edge[->,MRRGExtEdge] (in2);
	\node[              above right=.25 cm of in3] (an3) {}                 edge[->,MRRGExtEdge] (in3);
\end{tikzpicture}

}
	}
	\hfill
	\subfloat[
		A register for \iint@ = 3.
		Note the edges that connect from context 0 to context 1, 1 to 2 and 2 to 0.
	]{
		\label{fig:reg-mrrg}
		{

\begin{tikzpicture}
	\node[MRRGNode,                              ] (mo0) {\(\text{reg}_0^1\)};
	\node[MRRGNode,     right      =.20 cm of mo0] (mo1) {\(\text{reg}_1^1\)};
	\node[MRRGNode,     right      =.20 cm of mo1] (mo2) {\(\text{reg}_2^1\)};
	\node[MRRGNode,     below      =.25 cm of mo0] (ou0) {\(\text{out}_0\)}    edge[->,MRRGEdge] (mo0);
	\node[MRRGNode,     below      =.25 cm of mo1] (ou1) {\(\text{out}_1\)}    edge[->,MRRGEdge] (mo1) edge[<-,MRRGEdge] (mo0);
	\node[MRRGNode,     below      =.25 cm of mo2] (ou2) {\(\text{out}_2\)}    edge[->,MRRGEdge] (mo2) edge[<-,MRRGEdge] (mo1);
	\draw[->,MRRGEdge] (mo2.south west) -- ++(-0.3,-0.2) -- ([shift={(0.25,0.25)}]ou0.north east) -- (ou0.north east);
	\node[MRRGNode,     above      =.25 cm of mo0] (in0) {\(\text{in}_0\)}     edge[->,MRRGEdge    ] (mo0);
	\node[MRRGNode,     above      =.25 cm of mo1] (in1) {\(\text{in}_1\)}     edge[->,MRRGEdge    ] (mo1);
	\node[MRRGNode,     above      =.25 cm of mo2] (in2) {\(\text{in}_2\)}     edge[->,MRRGEdge    ] (mo2);
	\node[              below      =.25 cm of ou0] ()    {}                    edge[<-,MRRGExtEdge] (ou0);
	\node[              below      =.25 cm of ou1] ()    {}                    edge[<-,MRRGExtEdge] (ou1);
	\node[              below      =.25 cm of ou2] ()    {}                    edge[<-,MRRGExtEdge] (ou2);
	\node[              above      =.25 cm of in0] ()    {}                    edge[->,MRRGExtEdge] (in0);
	\node[              above      =.25 cm of in1] ()    {}                    edge[->,MRRGExtEdge] (in1);
	\node[              above      =.25 cm of in2] ()    {}                    edge[->,MRRGExtEdge] (in2);
\end{tikzpicture}

}
	}
	\end{minipage}

	\subfloat[
		A simple processing element for \iint@ = 2,
			consisting of a two-input latency-one ALU with one two-input mux on each input (``.m'' nodes),
			forming a crossbar from the input nodes to the FU nodes.
			The attachment points at the ``out'' nodes would typically
				connect to a register like \cref{fig:reg-mrrg}.
	]{
		\label{fig:pe-mrrg}
		{ 
\providecommand\figScaling{1.00}
	
\begin{tikzpicture}[
	on grid,
	node distance=1cm*\figScaling,
	every node/.append style={MRRGNode},
	every edge/.append style={->,MRRGEdge},
]
	\node[                      ] (out-t0) {\(\text{out}_0\)};
	\node[right=1 cm*\figScaling of out-t0] (out-t1) {\(\text{out}_1\)};
	\node[MRRGFUNode,above=of out-t0] (alu-t0) {\(\text{FU}_0^1\)}  edge (out-t1);
	\node[on grid=false, draw=none,below=.25 cm of out-t0] ()        {}                  edge[<-,MRRGExtEdge] (out-t0);

	\node[      left =of  alu-t0] (muxa-t0) {\(\text{a.m}_0\)};
	\node[below      =of muxa-t0] (outa-t0) {}                  edge ( alu-t0) edge[<-] (muxa-t0);
	\node[      left =of muxa-t0] (in0a-t0) {}                  edge (muxa-t0);
	\node[above left =of muxa-t0] (in1a-t0) {}                  edge (muxa-t0);

	\node[above      =of  alu-t0] (outb-t0) {}                  edge (alu-t0);
	\node[above      =of outb-t0] (muxb-t0) {\(\text{b.m}_0\)}  edge (outb-t0);
	\node[below left =of muxb-t0] (in0b-t0) {}                  edge (muxb-t0);
	\node[      left =of muxb-t0] (in1b-t0) {}                  edge (muxb-t0);

	\node[      left =of in1b-t0] (pin1-t0) {\(\text{in1}_0\)}  edge (in1a-t0) edge (in1b-t0);
	\node[below left =of pin1-t0] (pin0-t0) {\(\text{in0}_0\)}  edge (in0a-t0) edge (in0b-t0);
	\node[on grid=false, draw=none,left=.25 cm of pin1-t0] (){} edge[MRRGExtEdge] (pin1-t0);
	\node[on grid=false, draw=none,left=.25 cm of pin0-t0] (){} edge[MRRGExtEdge] (pin0-t0);

	\node[MRRGFUNode,above=of out-t1] (alu-t1) {\(\text{FU}_1^1\)}  edge (out-t0);
	\node[on grid=false, draw=none,below=.25 cm of out-t1] ()        {}                  edge[<-,MRRGExtEdge] (out-t1);

	\node[      right=of  alu-t1] (muxa-t1) {\(\text{a.m}_1\)};
	\node[below      =of muxa-t1] (outa-t1) {}                  edge ( alu-t1) edge[<-] (muxa-t1);
	\node[      right=of muxa-t1] (in0a-t1) {}                  edge (muxa-t1);
	\node[above right=of muxa-t1] (in1a-t1) {}                  edge (muxa-t1);

	\node[above      =of  alu-t1] (outb-t1) {}                  edge (alu-t1);
	\node[above      =of outb-t1] (muxb-t1) {\(\text{b.m}_1\)}  edge (outb-t1);
	\node[below right=of muxb-t1] (in0b-t1) {}                  edge (muxb-t1);
	\node[      right=of muxb-t1] (in1b-t1) {}                  edge (muxb-t1);

	\node[      right =of in1b-t1] (pin1-t1) {\(\text{in1}_1\)}  edge (in1a-t1) edge (in1b-t1);
	\node[below right =of pin1-t1] (pin0-t1) {\(\text{in0}_1\)}  edge (in0a-t1) edge (in0b-t1);
	\node[on grid=false, draw=none,right=.25 cm of pin1-t1] (){} edge[MRRGExtEdge] (pin1-t1);
	\node[on grid=false, draw=none,right=.25 cm of pin0-t1] (){} edge[MRRGExtEdge] (pin0-t1);
\end{tikzpicture}

}
	}
	\caption{%
		Example MRRG fragments.
		Subscripts indicate the value for \(t\),
			and superscripts indicate the latency, if any.
		Dotted arrows indicate points where these fragments would
			get attached to the rest of the MRRG.
		Nodes that may have an operation mapped to them have thick outlines.
		\label{fig:sample-mrrgs}
	}
\end{figure}

In \cref{fig:mux-mrrg} we have a single context MRRG fragment
	that is used to represent a multiplexer hardware element.
Because there is no latency associated with it,
	all connections between vertices are within the same context, 0.
In contrast, \cref{fig:reg-mrrg} is used to express
	a register of configurable latency in a multi-context CGRA
	and contains connections between contexts.

A typical implementation of a PE will contain one FU node
	connected to registers and input crossbars,
	with the whole structure duplicated for each context,
	such as \cref{fig:pe-mrrg}.
PEs will also typically have another FU node that provides the ``constant'' operation,
	corresponding to constants in the DFG.
The inputs and output nodes of PEs are then
	connected together according to the connectivity of the architecture.
To provide support for DFG inputs and outputs,
	typical CGRA models will also have
	several FU nodes that implement only these IO operations.

To keep mapping generic,
	even though the MRRG is formed from
	many stitched together graphs like those in
	\cref{fig:sample-mrrgs},
	\cgrame@ ignores this and only uses
	the connectivity of the flattened graph
	\cite{CGRAME-ILP}.

\subsection{Integer Linear Programming}
Integer linear programming (ILP) is a powerful and generic tool for
	specifying and solving combinatorial optimization problems.
An ILP consists of three parts:
	a set of integer variables,
	a set of inequality constraints on weighted sums of the variables,
	and optionally a cost function (another weighted sum) to choose the best solution.
Once these are specified, one of many free or commercial solvers can be used to find a solution.
A solver will always find the optimal solution given enough time,
	but in general ILP is NP-complete, and solve times may be lengthy.

\subsection{\cgrame@'s Existing Approach}
The entire problem of mapping to a CGRA can be described as
	taking a DFG that describes the computation and
	``finding it'' in the MRRG that describes the hardware.
This amounts to
	matching vertices in the DFG with MRRG vertices that support the operation,
	and edges in the DFG with paths in the MRRG.
Specifically, this is very similar to the directed subgraph homeomorphism problem~\cite{DirSubHomeo},
	except that paths starting at the same MRRG vertex may initially overlap,
	as the DFG is a hypergraph.
The existing approach,
	due to S.~A. Chin~\cite{CGRAME-ILP},
	directly encodes this problem in an ILP,
	with the general approach being:
	\emph{if an MRRG node is used by a particular DFG edge,
	then at least one of its fanout must be too}.
This maxim is applied to every node, for every DFG edge,
	resulting in \(\mathcal O(|E(\text{DFG})|\cdot |V(\text{MRRG})|)\) constraints.
Additionally,
	it requires that if two MRRG nodes are connected by an edge,
	at most one of them may have multiple fanins.
This is to implement a constraint to prevent self reinforcing loops in the mapping:
	a cycle of only routing nodes satisfies the maxim above.
While this approach generally results in a very large ILP,
	it is guaranteed to capture 100\% of the flexibility of the architecture.

\section{Connectivity-Based CGRA Mapping}\label{sec:the-idea}

\subsection{Connectivity}\label{sec:connectivity-intro}
In CGRAs, the small number of input ports on a ALU (usually 2)
	means that every processing element can have a fully-connected crossbar
	-- even in CGRAs with up to 8 PE inputs \cite{ADRES-long} (\adres@)
	or very flexible routing \cite{HyCUBE-intro} (\hycube@).
As a prototypical example,
	consider the orthogonally-connected CGRA in \cref{fig:simple-ortho-arch}.
A connection between adjacent processing elements is guaranteed.
If there is minimal need to negotiate routing,
	there should be a corresponding minimal need to
	explicitly model all the individual multiplexers and connections.
In this work
	we find this to be the case, even for \adres@ or \hycube@,
	though less so for our ``Clustered'' architecture
	(\cref{fig:ADRES-arch,fig:HyCUBE-arch,fig:clustered-arch}).
These architectures have
	similar computation capability,
	but differ primarily in how the PEs are connected to each other;
	HyCUBE is the most flexible, and Clustered and ADRES are less so.
Clustered has small links between islands of fully-connected PEs,
	while ADRES has folded-torus connectivity.

\begin{figure}
\small
\centering
\subfloat[]{\label{fig:simple-ortho-arch}\begin{tikzpicture}
	\tikzmath{ \maxX = 2; \maxY = 2; }
	\foreach \x in {0,...,\maxX} {\foreach \y in {0,...,\maxY} {
		\node [PEBlock,at={(\x,\y)}] (pe\x-\y) {\small PE};
	}}
	\foreach \x in {0,...,\maxX} {\foreach \y in {0,...,\maxY} {
		\tikzmath{ \nx = int(\x+1); \ny = int(\y+1); }
		\ifnum\nx>\maxX\else\draw[-] (pe\nx-\y) -- (pe\x-\y);\fi
		\ifnum\ny>\maxY\else\draw[-] (pe\x-\y) -- (pe\x-\ny);\fi
	}}
\end{tikzpicture}}
\hspace{1cm}
\subfloat[]{\label{fig:simple-ortho-arch-non-adj-mapping}\begin{tikzpicture}[outer sep=auto]
	\tikzmath{
		\maxX = 3-1;
		\maxY = 3-1;
		int \x, \y;
		for \x in {0,...,\maxX}{for \y in {0,...,\maxY}{
			let \used{\x,\y} = inuse;
		};};
		let \used{2,0} = unused;
		let \used{0,2} = unused;
	}

	\foreach \x in {0,...,\maxX} {
		\foreach \y in {0,...,\maxY} {
			\node [PEBlock,at={(\x,\y)},\used{\x,\y}] (pe\x-\y) {\small PE};
		}
	}

	\foreach \x in {0,...,\maxX} {\foreach \y in {0,...,\maxY} {
		\tikzmath{ \nx = int(\x+1); \ny = int(\y+1); }
		\ifnum\nx>\maxX\else\draw[-,unused] (pe\nx-\y) -- (pe\x-\y);\fi
		\ifnum\ny>\maxY\else\draw[-,unused] (pe\x-\y) -- (pe\x-\ny);\fi
	}}

	\node[DFGNode,at=(pe0-0)] (oa1) {o};
	\node[DFGNode,at=(pe1-0)] (a1)  {$+$} edge[DFGEdge,->] (oa1);
	\node[DFGNode,at=(pe2-1)] (m1)  {$*$};
	\draw[DFGEdge,->] (m1) -- (pe2-0.center) -- (a1);
	\node[DFGNode,at=(pe2-2)] (im1) {i} edge[DFGEdge,->] (m1);
	\node[DFGNode,at=(pe1-1)] (a2)  {$+$} edge[DFGEdge,->] (m1) edge[DFGEdge,->] (a1);
	\node[DFGNode,at=(pe0-1)] (ia2) {i} edge[DFGEdge,->] (a2);
	\node[DFGNode,at=(pe1-2)] (ia2) {i} edge[DFGEdge,->] (a2);
\end{tikzpicture}}

\caption{
	\protect\subref{fig:simple-ortho-arch}:
		A simple, orthogonally connected, CGRA.
		Any two adjacent PEs may be selected for ALU inputs,
			and each PE may simply route-through one input instead of using its ALU.
		And \protect\subref{fig:simple-ortho-arch-non-adj-mapping}:
		an example mapping of DFG that requires use of a route-through.
	}\label{fig:simple-ortho-arch-and-mapping}
\end{figure}
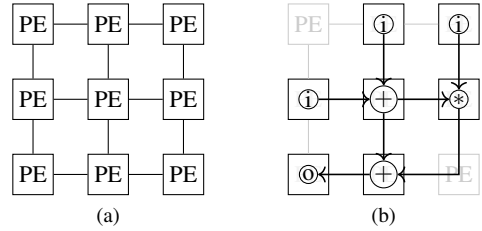

\tikzstyle{skipconRight} = [-,looseness=0.4,bend right=45]
\tikzstyle{skipconLeft} = [-,looseness=0.4,bend left=45]
\begin{figure}\centering
	\begin{tikzpicture}
	\small
	\tikzmath{\maxX = int(4-1); \maxY = int(4-1); }
	\foreach \x in {0,...,\maxX} {\foreach \y in {0,...,\maxY} {
		\node [PEBlock,at={(\x,\y)}] (pe\x-\y) {\small \ifnum\y=\maxX PE \else PE\fi};
	}}
	\foreach \y in {0,...,\maxY} {
		\node [draw, at={(-1.1,\y)}] (mem-\y) {\small MEM};
		\foreach \x in {0,...,\maxX} {
			\tikzmath{ \xbendright = \x < 2; }
			\ifnum\xbendright=1
				\draw (mem-\y) edge[-,black!60,looseness=0.4, bend right=45] (pe\x-\y);
			\else
				\draw (mem-\y) edge[-,black!60,looseness=0.3, bend left=45] (pe\x-\y);
			\fi
		}
	}
	\node[draw, rectangle, at={(\maxY/2,\maxY+1)}, minimum width=3cm] (rf) {Register File};
	\draw (pe0-\maxY) -- (rf.-150);
	\draw (pe1-\maxY) -- (rf.-120);
	\draw (pe2-\maxY) -- (rf.-60);
	\draw (pe3-\maxY) -- (rf.-30);
	\foreach \x in {0,...,\maxX} {\foreach \y in {0,...,\maxY} {
		\tikzmath{ \nx = int(\x+1); \ny = int(\y+1); }
		\ifnum\nx>\maxX\else\draw[-] (pe\nx-\y) -- (pe\x-\y);\fi
		\ifnum\ny>\maxY\else\draw[-] (pe\x-\y) -- (pe\x-\ny);\fi
	}}
	\foreach \x in {0,...,\maxX} {\foreach \y in {0,...,\maxY} {
		\tikzmath{\xbendright = int(mod(\x,2)); \ybendright = int(mod(\y,2)); }
		\tikzmath{\skipX = int(\x-2); \skipY = int(\y-2);}
		\ifnum\skipX<0\else
			\ifnum\y=\maxX\else 
				\ifnum\xbendright=1
					\draw (pe\x-\y) edge[skipconRight] (pe\skipX-\y);
				\else
					\draw (pe\x-\y) edge[skipconLeft] (pe\skipX-\y);
				\fi
			\fi
		\fi
		\ifnum\skipY<0\else
			\ifnum\ybendright=1
				\draw (pe\x-\y) edge[skipconRight] (pe\x-\skipY);
			\else
				\draw (pe\x-\y) edge[skipconLeft] (pe\x-\skipY);
			\fi
		\fi
	}}
\end{tikzpicture}
	\caption{\adres@ \cite{ADRES-long} architecture.
		An array of processing elements
		connected to vertically and horizontally adjacent \nei@s, and distance-two \nei@s.
		IO is done through the register file,
		and route-throughs are supported.
	}
	\label{fig:ADRES-arch}
\end{figure}
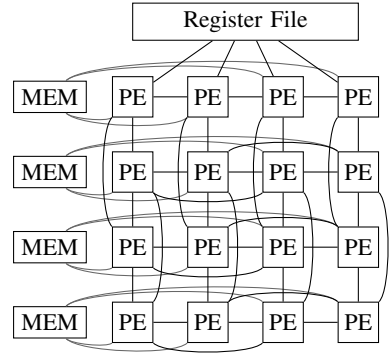

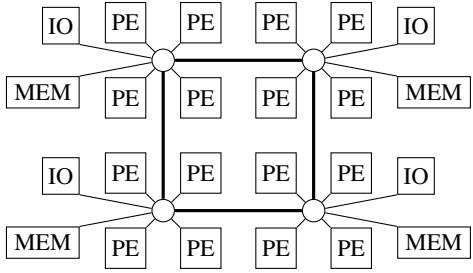
\begin{figure}\centering
	\begin{tikzpicture}
	\small
	\tikzmath{
		\maxX = 4-1;
		\maxY = 4-1;
		\maxXbX = int(\maxX/2);
		\maxXbY = int(\maxY/2);
	}
	\foreach \xid [evaluate=\xid as \x using \xid*2+.5] in {0,...,\maxXbX} {
		\foreach \yid [evaluate=\yid as \y using \yid*2+.5] in {0,...,\maxXbY} {
			\node [circle,at={(\x,\y)},draw] (xb\xid-\yid) {};
	}}
	\foreach \x in {0,...,\maxXbX} {\foreach \y in {0,...,\maxXbY} {
		\tikzmath{ \nx = int(\x+1); \ny = int(\y+1); }
		\ifnum\nx>\maxXbX\else\draw[-,very thick] (xb\nx-\y) -- (xb\x-\y);\fi
		\ifnum\ny>\maxXbY\else\draw[-,very thick] (xb\x-\y) -- (xb\x-\ny);\fi
	}}
	\foreach \x [evaluate=\x as \xbXID using int(\x/2)] in {0,...,\maxX} {
		\foreach \y [evaluate=\y as \xbYID using int(\y/2)] in {0,...,\maxY} {
			\node [PEBlock,at={(\x,\y)}] (pe\x-\y) {\small PE};
			\draw[-] (pe\x-\y) -- (xb\xbXID-\xbYID);
	}}
	\foreach \ID in {0,1} {
		\node[PEBlock, above left=.1 and 1 of xb0-\ID] (ioW-\ID) {IO};
		\draw (ioW-\ID.south east) edge[-] (xb0-\ID);
		\node[draw, rectangle, below left=.1 and 1 of xb0-\ID] (memW-\ID) {MEM};
		\draw (memW-\ID.north east) edge[-] (xb0-\ID);
		\node[PEBlock, above right=.1 and 1 of xb1-\ID] (ioE-\ID) {IO};
		\draw (ioE-\ID.south west) edge[-] (xb1-\ID);
		\node[draw, rectangle, below right=.1 and 1 of xb1-\ID] (memE-\ID) {MEM};
		\draw (memE-\ID.north west) edge[-] (xb1-\ID);
	}
\end{tikzpicture}
	\caption{A clustered architecture. Groups of 4 PEs connected to crossbars connected in a grid.
		PEs within a cluster are fully-connected,
			but the number of connections between clusters is limited to 1.
		Each cluster also has one memory port and one IO port.
	}
	\label{fig:clustered-arch}
\end{figure}

\begin{figure}\centering
	\begin{tikzpicture}
	\small
	\tikzmath{\maxX = int(4-1); \maxY = int(4-1); }
	\foreach \x in {0,...,\maxX} {\foreach \y in {0,...,\maxY} {
		\node [draw,circle,at={(\x,\y)}] (xb\x-\y) {};
		\node [PEBlock,at={(\x+0.5,\y+0.5)}] (pe\x-\y) {\small PE} edge[-] (xb\x-\y);
	}}
	\foreach \y in {0,...,\maxY} {
		\node [draw, rectangle, at={(-0.8,\y)}] (mem-\y) {\small MEM} edge[-] (xb0-\y);
		\node [PEBlock,at={(\maxX+1.25,\y)}] (ioW-\y) {IO} edge[-] (xb\maxX-\y);
	}
	\foreach \x in {0,...,\maxX} {
		\node [PEBlock,at={(\x,-0.6)}] (ioS-\x) {IO} edge[-] (xb\x-0);
		\node [PEBlock,at={(\x,\maxY+1.25)}] (ioN-\x) {IO} edge[-] (xb\x-\maxY);
	}
	\foreach \x in {0,...,\maxX} {\foreach \y in {0,...,\maxY} {
		\tikzmath{ \nx = int(\x+1); \ny = int(\y+1); }
		\ifnum\nx>\maxX\else\draw[-] (xb\nx-\y) -- (xb\x-\y);\fi
		\ifnum\ny>\maxY\else\draw[-] (xb\x-\y) -- (xb\x-\ny);\fi
	}}
\end{tikzpicture}
	\caption{\hycube@ \cite{HyCUBE-intro} Arch.
		A homogeneous array of PEs, each associated with a fully-connected crossbar which are connected in a grid.
	}
	\label{fig:HyCUBE-arch}
\end{figure}
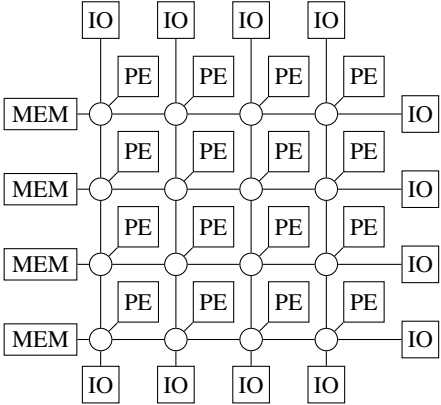

Further,
	given that we observe it is uncommon
	for the output of a PE to drive another PE that is far away,
	we present an ILP based around a more abstract principle:
	\emph{if an operation is mapped to a PE,
		then the fanin of the operation must be mapped to \nei@ing PEs}
		(\cref{con:fuf,con:mmo,con:fir,con:foiu} below).
Any choice of definition for a PE's \nei@ will work,
	but we opted for a heuristic method in this work:
	a symmetric breadth-first search in the MRRG.
To find the \nei@s of one FU node,
	a breadth-first search is started at that node,
	and after every expansion wave,
	the number of FUs discovered is compared to the target \numnei@.
If at least the target \numnei@ is found,
	then the search terminates and \nei@ are recorded.
This search is repeated for each FU node in the graph
	using the same target \numnei@,
	depending on the stage of the algorithm (details in \cref{sec:composition-intro}).
This is can be thought of as using a ``contracted'' MRRG:
	routing between ``\nei@ing'' FU nodes is eliminated,
	leaving only FUs connected to other FUs.

The target \numnei@ to search for before constructing the ILP
	must be carefully chosen.
Consider \cref{fig:simple-ortho-arch-non-adj-mapping}:
	there is no way to map the DFG to this CGRA without
	using another common feature of CGRAs: a PE route-through (see the lower-right PE in the figure).
Many CGRAs can convert a PE into a route-through,
	though others may provide features such as diagonal, torus and/or skip connections.
The \numnei@ that should be found
	in order to allow a DFG to be mapped is therefore architecture- \emph{and} DFG-dependent.
Effects of choosing particular \numnei@ are discussed in \cref{sec:exp:nn}.

\subsection{Paths}\label{sec:paths-intro}
For simple architectures like \cref{fig:simple-ortho-arch},
	it may be pointless to model routing congestion,
	but for more complicated architectures such as
	\cref{fig:clustered-arch,fig:HyCUBE-arch}
	we find it beneficial to model routing congestion.

From an intuitive view of the clustered architecture in \cref{fig:clustered-arch},
	it is obvious that in this CGRA
	there are limited connections between clusters.
However,
	deriving \nei@s from a breadth-first search is oblivious to this.
In fact, if we do not model routing, we find that
	for architectures like this, we must iterate through many placements before
	\todo{specific numbers?}
	finding one that can route.
To include routing in the ILP,
	we follow two principles:
	1) \emph{if an FU drives another FU, then a path through the MRRG from the driver to the driven must be chosen};
	and 2)
	\emph{two paths that are driven by different FUs cannot share a vertex}
	(\cref{con:prfe,con:pmex} below, respectively).

The potential downside of choosing paths from a finite set is
	that some of the combinatorial expressiveness of a graph-based device model can be lost.
Consider that between two FUs there may be a number of crossbars.
The number of paths between these two FUs is at least equal to the product of
	the widths of the crossbars.
The method used in this work is to find the 20
	cycle-less shortest paths between each pair of \nei@ing FU nodes,
	as this works for the 3 architectures under study.

\subsection{Composition}\label{sec:composition-intro}
As a first pass,
	a reasonable approach is to choose a sufficiently high \numnei@ to search for,
	identify a selected number of paths between each FU, and then try to solve
	a combined placement and routing ILP.
We found this to work well for trivially small architectures,
	but the large number of variables required for choosing paths
	results in an ILP with solve times greater than 20 minutes
	for most benchmarks on all 3 architectures in this work.

Alternatively,
	placement and routing can be completely split up
	by not modelling paths at all when finding a placement (``placement-only''),
	and then testing if a placement can be routed by a ``routing-only'' ILP that chooses paths
	for the connections required by the placement.
This approach has small ILPs for each stage,
	but the the placement-only ILP has no guidance as to routability.

To address this, one solution is to choose an ILP cost function
	for the placement
	that reflects how reliably a route can be found between two
	processing elements.
Another solution is to add a form of congestion modelling
	by adding the constraints for choosing paths,
	but relaxing the number of paths that may use the same vertex (i.e.~allowing shorts).
Implementing either of these approaches requires
	adding variables that model whether a pair of FUs are used
	(variables \(\edgevar\) below),
	and result in similarly sized ILPs.
In this work,
	we use the ``\relaxedPathConstraintPlacementName@''
	because it overall takes less time to discover a placement that will route
	-- even when compared to the cost-based approach
	with a cost function that heavily favours
	edges between FUs that are close together in the MRRG.
The quality of placements in either case is similar,
	but the addition of a cost function slows down
	the rate that solutions are discovered significantly.
We find that relaxing the vertex usage to at least 2 is effective,
	and that with this relaxation,
	we can reduce the number of paths for each connection to at least 3.

This \relaxedPathConstraintPlacementName@ produces an ILP big enough
	that the solver cannot consistently
	determine feasibility in less than \SI{200}{\milli\second}.
For example, the \numnei@ parameter may be too low to map the DFG,
	and the solver may take several tens of seconds to prove infeasibility.
However,
	we also aim to keep the \numnei@ as low as possible:
	a higher number results in an ILP with larger constraints and more solutions
	-- many of which will be unroutable, e.g.~requiring conflicting route-through usage.
Our strategy is to characterize an architecture to determine a schedule of \numnei@ to try,
	and use the
	the placement-only ILP as a test for a given \numnei@.
Its small model size makes it an low-runtime solution,
	and we find that even though this placement-only ILP does not model congestion,
	it is nearly 100\% predictive of
	whether a solution exists to the larger \relaxedPathConstraintPlacementName@ ILP.
Still,
	there are occasionally many solutions, with none routable,
	so we limit the number placements to 100 before moving on to the next \numnei@.
The proposed algorithm is the following:
\begin{lstlisting}[keywords={if,for,in,return}]
for nn in CGRA.(R"\pnei{}"R)CountSchedule():
	if placementOnly(nn).failure():
		continue;
	for pment in relaxedConstrPment(nn)[0:100]:
		routing = routingOnly(pment);
		if routing.success():
			return pment + routing;
return not_mappable;
\end{lstlisting}%

\section{ILP Formulation}\label{sec:the-details}

\label{sec:ilp-details}
The formulations discussed so far can be expressed as one ILP,
	with certain sets of constraints removed or relaxed to create the specific formulations.
The ILPs use the following definitions:

\begin{itemize}
	\item \(\text{FU} \subseteq V(\text{MRRG})\)
		: all MRRG nodes that can perform computation.
	\item \(\compatnodes(o \in \text{FU})\)
		: the set of MRRG FU nodes that \(o\) can be mapped to.
	\item \(\neighbours(u \in FU)\)
		: a set of MRRG FU nodes that are reachable from \(u\);
		\cref{sec:connectivity-intro}'s graph search results.
	\item \(\pathsfor(u \in \text{FU}, v \in \text{FU})\)
		: a set of paths through the MRRG from \(u\) to \(v\);
		\cref{sec:paths-intro}'s results.
\end{itemize}
The following variables describe the mapping:
\begin{itemize}
	\item \(\mathbf f_{ou}\) : is FU \(u\) used By DFG operation \(o\)?
	\item \(\mathbf p_{uvq}\) : is path number \(q\) used from FU \(u\) to \(v\)?
\end{itemize}
Also, a necessary intermediate variable class is used:
\begin{itemize}
	\item \(\mathbf e_{oupv}\) : is the DFG edge (\(o\),\(p\)) mapped to FUs \(u\) to \(v\)?
\end{itemize}

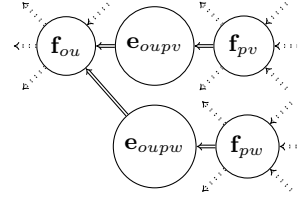
\begin{figure}
\small
\centering
\begin{tikzpicture}[node distance=0.25cm]
	\tikzstyle{fnode}=[draw, circle]
	\tikzstyle{enode}=[draw, circle]
	\tikzstyle{iedge}=[double,>=Implies] 
	\tikzstyle{eedge}=[iedge,dotted]

	\node (fou)   [fnode] {\(\mathbf f_{ou}\)};
	\node (eoupv) [enode, right=of fou  ] {\(\mathbf e_{oupv}\)} edge[->,iedge] (fou);
	\node (eoupw) [enode, below=of eoupv] {\(\mathbf e_{oupw}\)} edge[->,iedge] (fou);
	\node (fpv)   [fnode, right=of eoupv] {\(\mathbf f_{pv}\)}   edge[->,iedge] (eoupv);
	\node (fpw)   [fnode, right=of eoupw] {\(\mathbf f_{pw}\)}   edge[->,iedge] (eoupw);

	\draw (fou.north east)   edge[<-,eedge] ++( 0.3cm, 0.3cm);
	\draw (fou.south west)   edge[->,eedge] ++(-0.3cm,-0.3cm);
	\draw (fou.west)         edge[->,eedge] ++(-0.3cm, 0.0cm);
	\draw (fou.north west)   edge[->,eedge] ++(-0.3cm, 0.3cm);

	\draw (fpv.north east)   edge[<-,eedge] ++( 0.3cm, 0.3cm);
	\draw (fpv.east)         edge[<-,eedge] ++( 0.3cm, 0.0cm);
	\draw (fpv.south east)   edge[<-,eedge] ++( 0.3cm,-0.3cm);
	\draw (fpv.south west)   edge[->,eedge] ++(-0.3cm,-0.3cm);
	\draw (fpv.north west)   edge[->,eedge] ++(-0.3cm, 0.3cm);

	\draw (fpw.north east)   edge[<-,eedge] ++( 0.3cm, 0.3cm);
	\draw (fpw.east)         edge[<-,eedge] ++( 0.3cm, 0.0cm);
	\draw (fpw.south east)   edge[<-,eedge] ++( 0.3cm,-0.3cm);
	\draw (fpw.north west)   edge[->,eedge] ++(-0.3cm, 0.3cm);
	\draw (fpw.south west)   edge[->,eedge] ++(-0.3cm,-0.3cm);
\end{tikzpicture}
\caption{Visualisation of the relation between edge and FU variables. Arrows indicate implication.}\label{fig:fu-edge-var-visual}
\end{figure}

\cref{fig:fu-edge-var-visual} visualizes the relationship between the
	FU-is-used variables (\(\mathbf f\)) and edge-is-used variables (\(\mathbf e\)).
The placement-only ILP uses all constraints described below, except
	\ref{con:prfe} and \ref{con:pmex}.
The routing-only ILP only uses
	constraints \ref{con:prfe} and \ref{con:pmex},
	with \(\mathbf e\) variables set to fixed values.
Finally,
	the \relaxedPathConstraintPlacementName@ ILP uses all constraints as stated,
	except \ref{con:pmex},
	which is relaxed as described below.


\ILPConstraint{Functional Unit Exclusivity}\label{con:fuf}
This constraint ensures that each physical functional unit is not occupied by multiple DFG vertices.
\begin{gather*}
	\forall u \in \text{FU} \quad
	\sum_{o \in \text{DFG}} \mathbf f_{ou} \leq 1
\end{gather*}

\ILPConstraint{Must Map Ops}\label{con:mmo}
Requires that a mapping be found.
Due to later constraints, it is sufficient to only apply this constraint for output nodes,
	specifically, a set DFG vertices whose combined fanin cones cover the entire DFG.

Also, if this constraint is relaxed to be greater-than-or-equal-to 1 (instead of exactly equal to 1),
	then the ILP supports duplication of operations, i.e.~re-computation.
The effects of this are not explored in this work.
We suspect it may be useful for certain DFGs and architectures,
	but that it also may hinder finding useful mappings by
	increasing the number of unwanted mappings
	(such as filling the architecture with redundant computation).
\begin{gather*}
	\forall o \in \graphoutputs(\text{DFG}) \quad
	\sum_{u \in \compatnodes(o)} \mathbf f_{ou} = 1 \quad (\text{or} \geq 1)
\end{gather*}

\ILPConstraint{Fanin Required}\label{con:fir}
This constraint encodes the notion that if an operation is mapped to an FU,
	then each DFG fanin of the operation must be mapped to a \nei@ing FU,
	i.e. \(\mathbf f_{pv} \Rightarrow \exists u : \mathbf e_{oupv}\) for each fanin \(o\).
This constraint ensures that all data that are needed by FU \(v\) will arrive.
\begin{gather*}
	\forall (o,p) \in \text{DFG} \ \forall v \in \compatnodes(p) \\
	\mathbf f_{pv} \leq \sum_{u \in \compatnodes(o) \text{ if } v \in \neighbours(u) } \mathbf e_{oupv}
\end{gather*}

\ILPConstraint{Fanout Implies Usage}\label{con:foiu}
If an edge variable originates at a given FU, then that FU must be in use by the fanin operation,
	i.e. \(\mathbf e_{oupv} \Rightarrow \mathbf f_{ou}\).
\begin{gather*}
	\forall (o,p) \in \text{DFG} \ \forall u \in \compatnodes(o) \ \forall v \in \compatnodes(p) \cap \neighbours(u) \\
	\mathbf e_{oupv} \leq \mathbf f_{ou}
\end{gather*}

\ILPConstraint{Path Required for an Edge}\label{con:prfe}
Simply,
	if an edge is in use, then at least one path corresponding to it must be in use,
	i.e. \(\mathbf e_{oupv} \Rightarrow \exists q \mathop{:} \mathbf p_{uvq}\).
\begin{gather*}
	\forall (o,p) \in \text{DFG} \ \forall u \in \compatnodes(o) \ \forall v \in \compatnodes(p) \cap \neighbours(u) \\
	\mathbf e_{oupv} \leq \sum_{q \in \pathsfor(u,v)} \mathbf p_{uvq}
\end{gather*}

\ILPConstraint{Paths are Mutually Exclusive if Driven by Different FUs}\label{con:pmex}
If a path through the MRRG is mapped to a DFG edge,
	then it electrically cannot overlap with another path,
	unless both paths are driven by the same physical FU.
\begin{gather*}
	\forall u,v,w,x \in \text{FU} \ \forall q \in \pathsfor(u,v) \\
		\forall z \in \{ \pathsfor(w,x) \mathop{:} \lnot \compatpaths(u,q,w,z) \} \\
	\mathbf p_{uvq} +
		\mathbf p_{wxz} \leq 1
\end{gather*}

Where
	\(\compatpaths(u,q,w,z) = ( q \cap z \questeq \varnothing ) \lor ( u \questeq w )\),
	i.e. do the paths not overlap, or are they driven by the same FU node.
By relaxing this constraint to be less than or equal to some integer greater than 1,
	an ILP that allows routing overuse is created.
Also, the constraint as-specified will produce duplicate constraints,
	which are detected and not added to the ILP.

For ILPs that model congestion or routing,
	the variables that are required by this constraint category
	will easily dominate the number of variables attributed to other categories.
More precisely,
	the number of \(\mathbf p\) variables is linear with respect to
	the CGRA size, \numnei@ used, and number of paths used
	-- \(\mathcal O(|\text{FU}|\cdot N\cdot P)\),
	so the number possible conflicts is in
	\(\mathcal O(|\text{FU}|^2 \cdot N^2 \cdot P^2)\).

\ILPConstraint{With a Generic Cost Function}\label{con:cfunc}
With variables representing each aspect of a mapping,
	various cost functions can be specified.
Using the first summation below, the coefficients \(k_{ou}\) can specify that
	a certain FU node \(u\) for DFG vertex \(o\) is preferred over other FU nodes,
	such as to encourage using a certain portion of the CGRA.
The \spellCentre@ summation can be used to select certain FU-FU connections
	to be preferentially chosen for particular DFG edges,
	or, that using certain pairs of FU nodes is preferred over others.
For example, if coefficients \(l_{oupv}\) are set to the taxicab distance between the FUs \(u\) and \(v\),
	then the half-perimeter bounding box is minimized.
The bottom summation can be used to encourage
	the choice of particular paths,
	such as by setting \(m_{uvg}\) to the estimated power consumption of using that path.
\begin{gather*}
	\mathcal L =
	\sum_{o \in \text{DFG}\vphantom{(o)}} \ \sum_{u \in \compatnodes(o)}
		k_{ou}\mathbf f_{ou} \\
	+ \sum_{(o,p) \in \text{DFG}} \ \sum_{u \in \compatnodes(o)} \ \sum_{v \in \compatnodes(p) \cap \neighbours(u)}
		l_{oupv}\mathbf e_{oupv} \\
	+ \sum_{u,v \in \text{FU}\vphantom{(u,v)}} \ \sum_{q \in \pathsfor(u,v)}
		m_{uvq}\mathbf p_{uvq}
\end{gather*}


\section{Experimental Results}\label{sec:exp-res}
\subsection{Setup \& Overview}\label{sec:exp:setup}
\label{sec:exp:nn}

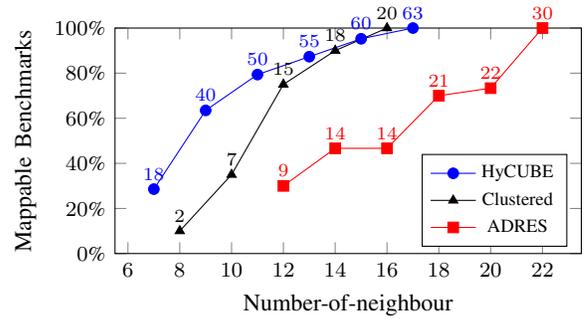
\begin{figure}\centering
\begin{tikzpicture}
	\begin{axis}[
		xlabel={\Numnei@},
		y=3cm,
		ymin=0,
		ymax=1,
		ylabel={Mappable Benchmarks},
		ytick distance=0.2,
		yticklabel={\pgfmathparse{\tick*100}\pgfmathprintnumber\pgfmathresult\%},
		legend entries={HyCUBE,Clustered,ADRES},
		legend pos=south east,
		legend style={font=\scriptsize},
		unbounded coords=discard, 
		filter discard warning=false, 
		nodes near coords={\scriptsize\pgfmathprintnumber\pgfplotspointmeta},
	]
		\addplot [blue, mark=*] table [x=NN,
			y expr=\thisrow{hycube-4x4-mapped}/\thisrow{hycube-4x4-mappable},
			point meta=\thisrow{hycube-4x4-mapped}
		] {data/number-of-neighbour-stats.txt};
		\addplot [black, mark=triangle*] table [x=NN,
			y expr=\thisrow{clustered-2x2-mapped}/\thisrow{clustered-2x2-mappable},
			point meta=\thisrow{clustered-2x2-mapped}
		] {data/number-of-neighbour-stats.txt};
		\addplot [red, mark=square*] table [x=NN,
			y expr=\thisrow{adres-4x4-mapped}/\thisrow{adres-4x4-mappable},
			point meta=\thisrow{adres-4x4-mapped}
		] {data/number-of-neighbour-stats.txt};
	\end{axis}
\end{tikzpicture}
\caption{
	Number of benchmarks that will map for a given \numnei@,
		normalized to the number of benchmarks that will map to that architecture.
	Different values of \iint@ are presented as the same architecture.
}\label{fig:nn-mapfrac}
\end{figure}

To simulate various sizes of architecture,
	and to demonstrate variability in runtime,
	we test each architecture with 1, 2 and 3 CGRA contexts.
We use the set of benchmarks \cite{CGRAME-ILP} used
	to characterize the existing \cgrame@ approach,
	plus some additional computation applications,
	and a fast-Fourier-transform benchmark from MiBench \cite{MiBench}.
\ifCLASSOPTIONpeerreview\else
We are limited by our range of benchmarks because our DFG generator
	does not support conditional statements
	\cite{CGRAME-intro}.
\fi
Each benchmark-architecture-\iint@ combination is
	run 6 times using different ILP solver seeds for robust runtime measurements,
	and with a limit of 7.5 hours -- near the length of a typical workday.

The algorithm from \cref{sec:composition-intro} is used,
	with the schedules for \numnei@ determined empirically for each architecture,
	and presented as the horizontal values of the curves in \cref{fig:nn-mapfrac}.
We start with the minimum value, and increment by 2 until the maximum.
With these schedules,
	this approach provides the same feasibility results
	as the existing approach,
	whenever the existing approach can
	provide a decision within the time limit.
\cref{fig:nn-mapfrac}
	also shows the utility of starting at a small \numnei@ and increasing slowly:
	many benchmarks can be mapped without reaching the maximum value.
The differences between architectures is also notable.
\hycube@
	will map many benchmarks at a low \numnei@
	due to it flexible routing,
	however some benchmarks require long-distance connections.
Clustered
	can map suddenly many more benchmarks
	once inter-cluster FUs are consider \nei@s,
	but soon runs into the limits of the architecture.
\adres@
	is much more homogeneous than the other two architectures,
	resulting in a more linear trend.
We also suggest a generic schedule for new architectures:
	start at 4 with step 2, ending at 24.

The time taken to run the n-shortest-paths algorithm to find paths has been
	removed from the runtime of our approach by caching all data in memory before commencing the timer.
Filling this cache with 20 paths between every pair of FU nodes (excessive, but guaranteed to cover all possibilities)
	using a single CPU thread takes at most 5.4 seconds for the largest architecture tested,
	\hycube@ with \iint@ = 3.
Each run of the n-shortest-paths algorithm takes very little time and is completely independent,
	so parallelism could trivially be used to reduce this time.
And, the paths discovered do not change for a given architecture and \iint@,
	so they could be cached to disk.

The ILP solver used for both is the one provided by Gurobi Software \cite{gurobi-solver},
	and experiments were performed using up to 4 threads on Intel\textregistered{} Xeon\textregistered{} Gold 6148 Processors.

\subsection{Comparison \& Discussion}\label{sec:exp:discussion}%
\todo{semi visible stacked that showing cached compute time?}%
\todo{Numbers for/Impact of test placements?}%
\todo{mention that num route paths doesn't change route runtime}%
\todo{scatter plots showing/not showing specific correlations mentioned}%
\todo{sort benchmark data by minII}%
\todo{mention cap is unmappable}%
Twenty-eight benchmarks over 3 architectures times 3 \iint@ values is too many data-points to present directly,
	so summaries of runtime data are presented instead,
	in \cref{tab:bench-runtime-comp,tab:arch-runtime-comp}.
We take geometric means across
	architecture-II or benchmark axes
	and compute the relative speedup.
Maximum runtimes of this work are also presented.
At the left of \cref{tab:bench-runtime-comp}, there is the benchmark that is selected for each row,
	presented along with the number of vertices in the benchmark DFG.
To compute column 3,
	the average runtime over the seeds is determined for each architecture and II,
	and the geometric mean of the averages is taken.
The computation for column 4 is exactly the same, except runtimes for this work are used.
The speedup column is simply the value of column 3 divided by column 4.
Column 6 is similar to column 4, except the maximum of the runtime averages is taken, instead of geometric mean.
At the far right is the median \numnei@ required to
	map this benchmark on the architecture and II variants.
\cref{tab:arch-runtime-comp} is similar, except the statistics are
	taken across all benchmarks while fixing architecture and \iint@.

\pgfplotstableset{arch bench formatting/.style={
	column type=l,
	sort=true,
	sort key={heur-rt-geo},
	sort cmp=float <,
	numeric type,
	fixed,
	fixed zerofill=true,
	precision=1,
	skip 0.=true,
	columns/big-rt-geo/.style={column name=Existing \cite{CGRAME-ILP}, dec sep align},
	columns/geo-geo-spup/.style={column name=Speedup, dec sep align},
}}

\begin{table*}\centering
\caption{Geometric means of six-seed architecture arithmetic means, and maximum runtimes, by benchmark}\label{tab:bench-runtime-comp}
\begin{filecontents*}{data/bench-time-table.txt}
          id	bench          	  bench-size	    min-iint	min-avg-map-nn	med-avg-map-nn	avg-avg-map-nn	max-avg-map-nn	  big-rt-min	  big-rt-geo	  big-rt-med	  big-rt-avg	  big-rt-max	big-num-mappped	 heur-rt-min	 heur-rt-geo	 heur-rt-med	 heur-rt-avg	 heur-rt-max	heur-num-mapped	avg-avg-spup	geo-geo-spup	min-max-spup	max-max-spup
           0	accum2         	          18	           1	     7.00000	     9.00000	    12.60000	    22.00000	    76.01540	  6971.45709	 27000.00000	 18196.77156	 27048.70000	           3	     1.29744	    20.19895	    18.72650	    55.12775	   242.94928	           5	   330.08367	   345.13957	     0.31289	   111.33476
           1	accumulate     	          18	           1	     7.00000	     9.00000	    12.60000	    22.00000	   371.80500	 11046.88733	 27000.00000	 18725.27300	 27000.10000	           3	     1.33343	    20.65107	    22.21821	    54.28972	   235.57176	           5	   344.91379	   534.93051	     1.57831	   114.61518
           2	add-10         	          20	           1	     7.33333	     9.00000	    10.26667	    14.00000	     0.41692	    93.93084	    29.46007	  7494.73558	 27000.10000	           3	     0.12842	     4.10789	     9.06676	    32.18987	   175.72426	           5	   232.82898	    22.86593	     0.00237	   153.65039
           3	add-14         	          28	           2	     9.00000	     9.00000	    10.00000	    12.00000	     0.73866	    17.51696	     2.58110	  6001.92357	 27000.10000	           0	     0.18178	     2.35950	     0.46718	    45.19010	   412.16869	           3	   132.81501	     7.42401	     0.00179	    65.50740
           4	add-16         	          32	           2	     9.33333	    10.00000	    10.44444	    12.00000	     0.58770	    19.31183	     3.07133	  6002.23807	 27000.20000	           0	     0.18908	     2.30996	     0.45090	    67.04490	   583.92272	           3	    89.52565	     8.36023	     0.00101	    46.23934
           5	cap            	          24	         nan	         nan	         nan	         nan	         nan	     0.11100	     0.62912	     0.56089	     0.98666	     4.43953	           0	     0.12371	     0.27808	     0.28651	     0.30656	     0.59183	           0	     3.21843	     2.26238	     0.18756	     7.50132
           6	conv2          	          16	           1	     7.00000	    10.00000	    10.19048	    12.33333	     3.62602	  3091.29031	 27000.00000	 18031.90575	 27001.50000	           3	     1.15949	    13.25340	    19.57724	    21.76852	    77.75046	           7	   828.34779	   233.24506	     0.04664	   347.28413
           7	conv3          	          24	           1	     9.00000	    10.33333	    10.60000	    12.33333	    53.93550	  5469.71335	 27000.00000	 18124.89705	 27000.10000	           3	     3.02440	    31.64069	    19.32713	    66.48128	   282.84887	           5	   272.63158	   172.86959	     0.19069	    95.45769
           8	cos-4          	          21	           1	    15.40000	    15.70000	    15.70000	    16.00000	     0.61410	  2436.71400	 27000.01667	 18560.55912	 27010.70000	           1	     0.20541	   119.93524	   236.50768	   253.67140	   583.80756	           2	    73.16772	    20.31691	     0.00105	    46.26644
           9	cosh-4         	          21	           1	    15.00000	    15.00000	    15.00000	    15.00000	     0.59151	  2449.76208	 27000.00000	 18463.85420	 27002.20000	           1	     0.16935	   119.56203	   234.18029	   259.41938	   561.68002	           1	    71.17377	    20.48947	     0.00105	    48.07399
          10	exponential-4  	          13	           1	     9.66667	    12.50000	    12.03333	    14.00000	     0.47572	   509.12073	  1538.99735	  4519.30629	 27000.00000	           7	     0.14257	    35.57076	    66.28744	    81.09686	   266.02603	           5	    55.72727	    14.31290	     0.00179	   101.49383
          11	exponential-5  	          19	           1	    12.00000	    15.40000	    14.80000	    17.00000	     0.65825	  2295.80857	 27000.00000	 18311.64260	 27000.50000	           1	     0.17914	   135.35726	   338.19111	   291.89653	   522.44725	           3	    62.73333	    16.96110	     0.00126	    51.68082
          12	exponential-6  	          26	         nan	         nan	         nan	         nan	         nan	     1.07377	  1270.51797	 27000.01667	 18001.41975	 27011.80000	           0	     0.20847	    39.54689	   401.96212	   311.38414	   903.61535	           0	    57.81097	    32.12688	     0.00119	    29.89303
          13	FFT            	          38	         nan	         nan	         nan	         nan	         nan	     2.23967	   242.12198	     7.50147	 12004.82929	 27051.50000	           0	     0.26689	    11.44028	     0.95151	   343.10499	  1680.26033	           0	    34.98879	    21.16399	     0.00133	    16.09959
          14	mac            	          11	           1	     7.00000	    10.00000	    12.42857	    22.00000	     2.83828	   401.37601	   446.06772	  6613.23168	 27000.00000	           7	     0.47706	     5.26611	     5.30635	    12.58151	    56.71588	           7	   525.63119	    76.21868	     0.05004	   476.05712
          15	mac2           	          24	           1	     9.00000	    13.50000	    14.50000	    22.00000	   521.38000	 15372.43994	 27000.03333	 19009.45580	 27002.70000	           4	     4.46707	    52.69671	    47.70783	   131.57490	   461.79796	           4	   144.47631	   291.71536	     1.12902	    58.47297
          16	matrixmultiply 	          17	           1	     7.00000	     9.50000	    12.16667	    22.00000	     8.00139	  1370.96202	  3029.62500	  9528.52133	 27000.10000	           7	     0.75284	    10.01209	     9.24148	    30.99687	   179.00428	           6	   307.40269	   136.93066	     0.04470	   150.83494
          17	multiply-10    	          20	           1	     7.33333	     9.00000	     9.33333	    12.00000	     0.31755	    89.26726	    36.87326	  7301.71804	 27000.00000	           3	     0.13373	     4.30984	     9.66399	    35.58011	   230.30089	           4	   205.21907	    20.71245	     0.00138	   117.23793
          18	multiply-14    	          28	           2	     9.00000	     9.00000	    10.00000	    12.00000	     0.48704	    16.35266	     2.58265	  6001.98780	 27006.30000	           0	     0.14851	     2.21349	     0.41625	    40.94562	   374.45504	           3	   146.58437	     7.38772	     0.00130	    72.12161
          19	multiply-16    	          32	           2	     9.33333	    10.00000	    10.44444	    12.00000	     0.65736	    18.85169	     3.57783	  6002.22600	 27000.10000	           0	     0.16254	     2.19025	     0.42422	    59.78766	   591.35072	           3	   100.39238	     8.60709	     0.00111	    45.65835
          20	mults2         	          25	           1	     9.00000	    15.00000	    15.13333	    22.00000	  1070.48000	 19866.93946	 27000.01667	 21891.86398	 27005.50000	           3	     9.22798	    86.20898	    96.69713	   148.88189	   429.17329	           5	   147.04182	   230.45093	     2.49428	    62.92447
          21	nomem1         	           6	           1	     7.00000	     8.00000	    11.85714	    22.00000	     0.31319	   117.44632	    64.05319	  6189.70162	 27000.00000	           7	     0.13976	     2.29141	     2.00206	     6.45230	    27.10502	           7	   959.30123	    51.25495	     0.01155	   996.12537
          22	simple         	          12	           1	     7.00000	    10.00000	     9.85714	    12.00000	     3.48407	  2674.14043	 27000.00000	 18020.86446	 27000.10000	           3	     0.65286	     6.98170	     7.71857	    12.37579	    55.16947	           7	  1456.13833	   383.02148	     0.06315	   489.40295
          23	simple2        	          12	           1	     7.00000	    10.00000	     9.85714	    12.00000	     2.18389	  2551.53944	 27000.00000	 18018.63660	 27001.70000	           3	     0.53384	     7.05222	     7.79872	    12.41894	    53.49656	           7	  1450.90009	   361.80649	     0.04082	   504.73712
          24	sum            	           7	           1	     7.00000	     9.00000	    12.14286	    22.00000	     0.43573	    82.01531	    38.46221	  6104.06856	 27000.00000	           7	     0.25158	     3.79804	     4.62873	    11.16317	    50.00923	           7	   546.80430	    21.59412	     0.00871	   539.90035
          25	taylor-series-4	          15	           1	     8.00000	    12.00000	    11.33333	    13.00000	     0.30828	  1065.33152	 10123.40750	 10699.26211	 27000.10000	           5	     0.15135	    47.36512	    82.52771	   105.36649	   291.27634	           5	   101.54331	    22.49190	     0.00106	    92.69582
          26	long-chain     	          35	           2	    12.50000	    12.75000	    12.75000	    13.00000	     1.08514	    29.17813	     7.24154	  6004.23888	 27004.90000	           0	     0.23665	    10.74496	     0.53515	   296.22522	  1368.11818	           2	    20.26917	     2.71552	     0.00079	    19.73872
          27	weighted-sum   	          32	           2	     9.33333	     9.66667	     9.66667	    10.00000	     0.55661	    20.90319	     4.56581	  6002.55544	 27000.00000	           0	     0.16802	     2.34084	     0.43627	    70.54055	   595.68003	           2	    85.09369	     8.92979	     0.00093	    45.32635
\end{filecontents*}
\pgfplotstabletypeset[
	columns={[index]1,bench-size,big-rt-geo,heur-rt-geo,geo-geo-spup,heur-rt-max,med-avg-map-nn},
	my formatting,
	arch bench formatting,
	display columns/0/.style={column name=Benchmark Name, string type},
	columns/bench-size/.style={column name=DFG Size, int detect, column type=r, clear infinite},
	columns/heur-rt-geo/.style={column name=This Work, dec sep align, clear infinite},
	columns/heur-rt-max/.style={column name=This Work Max., dec sep align, clear infinite},
	columns/med-avg-map-nn/.style={column name=Median NN, fixed, precision=0, column type=r, clear infinite},
]{data/bench-time-table.txt}
\end{table*}

\begin{table}\centering
\caption{Geometric means of six-seed benchmark arithmetic means, and maximum runtimes, by architecture}\label{tab:arch-runtime-comp}
\begin{filecontents*}{data/arch-time-table.txt}
id	arch                     	II	big-rt-min	big-rt-geo	big-rt-med	big-rt-avg	big-rt-max	big-num-mappped	heur-rt-min	heur-rt-geo	heur-rt-med	heur-rt-avg	heur-rt-max	heur-num-mapped	avg-avg-spup	geo-geo-spup	min-max-spup	max-max-spup
0	adres-4x4                	1	     0.11100	     4.41826	     1.16330	   415.05309	 15285.10000	0	     0.12371	     1.45600	     0.27905	    59.01108	   461.79796	2	     7.03348	     3.03453	     0.00024	    33.09911
1	adres-4x4                	2	     0.36450	   310.16828	   985.54976	 11851.50052	 27001.30000	0	     0.23306	     7.12853	     7.21615	    87.88702	   728.04123	2	   134.84926	    43.51085	     0.00050	    37.08760
2	adres-4x4                	3	     0.49212	  1303.74390	 27000.00000	 15613.06034	 27021.90000	0	     0.32177	    15.60491	    27.78452	   108.40129	  1680.26033	2	   144.03020	    83.54704	     0.00029	    16.08197
3	clustered-2x2            	1	     0.16961	     7.76520	     2.00960	  1542.39496	 27000.00000	0	     0.12842	     5.19858	     1.69757	   107.02533	   583.80756	2	    14.41149	     1.49372	     0.00029	    46.24812
4	clustered-2x2            	2	     0.38305	   871.16490	 27000.00000	 15968.68805	 27001.20000	0	     0.22761	    24.68436	    20.37922	   114.32842	   723.89690	2	   139.67383	    35.29219	     0.00053	    37.29979
5	clustered-2x2            	3	     0.72351	  2902.46996	 27000.00000	 18993.18071	 27051.50000	0	     0.36207	    72.13623	    56.33387	   160.43949	   876.77005	2	   118.38220	    40.23596	     0.00083	    30.85359
6	hycube-4x4               	1	     0.37297	    79.56654	   104.79024	  1212.21061	 26191.30000	0	     0.16499	     4.28471	     1.87049	    67.40047	   522.44725	2	    17.98519	    18.56987	     0.00071	    50.13195
7	hycube-4x4               	2	     1.41598	  8422.27472	 27000.00833	 19365.46186	 27011.80000	0	     0.31462	    13.45727	     7.24294	   101.25691	  1368.11818	2	   191.25077	   625.85304	     0.00103	    19.74376
8	hycube-4x4               	3	     3.19231	 12788.05962	 27000.01667	 21054.80962	 27048.70000	0	     0.48926	    23.17878	    18.71694	   112.84879	  1023.94686	2	   186.57541	   551.71421	     0.00312	    26.41612
\end{filecontents*}
\pgfplotstabletypeset[
	columns={[index]1,II,big-rt-geo,heur-rt-geo,geo-geo-spup,heur-rt-max},
	my formatting,
	arch bench formatting,
	display columns/0/.style={column name=Arch. Name, string type},
	columns/II/.style={column name=II, int detect, column type=c},
	columns/heur-rt-geo/.style={column name=Time, dec sep align},
	columns/heur-rt-max/.style={column name=Max. Time, dec sep align},
	every head row/.style={
		before row={\toprule & & & & \multicolumn{6}{|c|}{This Work}\\},
		after row=\midrule,
	},
]{data/arch-time-table.txt}
\end{table}

Looking at \cref{tab:bench-runtime-comp},
	it is clear that certain benchmarks tend to produce long runtimes in the existing approach.
However, there is no clear correlation between DFG size or complexity,
	with one of the largest and most complicated benchmarks (FFT) having
	a short average runtime compared to some smaller, simpler DFGs (eg. conv2, accumulate).
This could be attributed to the solver quickly determining that
	there are not enough FUs to map FFT on \iint@ = 1 variants, which is the case,
	however conv2 and accumulate, which are able to map with \iint@ = 1,
	cause timeouts on \iint@ > 1 architectures.

Looking at the average runtime for this work in \cref{tab:bench-runtime-comp},
	we see that there is a again not much of a correlation between with DFG size.
However,
	looking at the median \numnei@ required to map the benchmarks
	there is a correlation:
as the number of path exclusivity constraints increases with the square of the \numnei@,
	benchmark that requires a high number of fanout (exponential-*,cosh-4,cosh-4) or
	onnections to distant FU nodes (FFT, long-chain) will take more time to map.

Here, we observe a more straightforward pattern in the existing approach:
	an increasing \iint@ (which essentially acts as a multiplier on MRRG size)
	results in increasing runtime.
This is also true for this work,
	as a higher \iint@ implies more FU nodes which implies again that more path variables must exist.

When considering each benchmark-architecture-II experiment individually,
	the speedup in geometric mean of all data-points is \geomeanSpeedup@\(\times\).
The arithmetic mean of speedups is \arithMeanSpeedup@\(\times\), and the median speedup is \medianSpeedup@\(\times\).
The 90th percentile average runtime for the new approach is \heurTimeNinteyP@ seconds.

For problems that
	the existing approach can provide an mapping (or infeasible result) within the time limit,
	this approach matches for all all but 17 (6.7\%).
All non-matching results are cases where
	this approach is not able to produce a mapping and the existing approach can.
Fifteen of the non-matching results are for \iint@ = 1,
	suggesting that this approach has some trouble with highly constrained problems.
Also, for \adres@ and \hycube@,
	the \relaxedPathConstraintPlacementName@ ILP produces the same
	feasible/infeasible result as the existing approach.
For architecture exploration,
	this may be sufficient
	and if used, would have a 244\(\times\) speedup over the existing approach.
We believe that the reason that the \relaxedPathConstraintPlacementName@ ILP
	finds a solution on the \clustered@ architecture is that allowing shorts
	does not model the limited connectivity between clusters.

When there is a solution to the \relaxedPathConstraintPlacementName@ ILP,
	but it is not possible to map (ie. \clustered@)
	the runtime is entirely attributable to
	constructing and solving the 100 routing-only ILPs per \numnei@,
	which may take up to 3 seconds each.
This also occurs to a lesser degree on
	the other architectures when a benchmark is only barely
	not routable at some NN.

\begin{table}\centering
	\caption{Sample MRRG sizes \& corresponding ILP sizes after presolving for taylor-series-4 benchmark, with \iint@ = 2}
	\label{tab:mrrg-size-stats}
	\begin{tabularx}{\linewidth}{ X *5c}
		\toprule
		\multicolumn{2}{c}{} & \multicolumn{2}{c}{No. Constraints} & \multicolumn{2}{c}{No. Variables} \\ 
		\midrule
		Arch. & MRRG Size & Existing & This Work & Existing & This Work \\ 
		\midrule
		\adres@   & 2320 & 17567 & 19159 & 18206 & 27472 \\ 
		\multicolumn{2}{l}{\quad Placement-only} & & 423 & & 336 \\ 
		\multicolumn{2}{l}{\quad Routing-only} & & 8525 & & 410 \\ 
		Clustered & 4488 & 22409 & 19247 & 22002 & 25728 \\ 
		\multicolumn{2}{l}{\quad Placement-only} & & 511 & & 384 \\ 
		\multicolumn{2}{l}{\quad Routing-only} & & 6931 & & 412 \\ 
		\hycube@  & 5056 & 52995 & 12947 & 42208 & 17896 \\ 
		\multicolumn{2}{l}{\quad Placement-only} & & 599 & & 432 \\ 
		\multicolumn{2}{l}{\quad Routing-only} & & 4421 & & 412 \\ 
		\bottomrule
	\end{tabularx}
\end{table}

Finally,
	\cref{tab:mrrg-size-stats}
	presents some sample MRRGs' vertex set sizes
	with corresponding ILP statistics.
The largest ILP used by this work
	is the \relaxedPathConstraintPlacementName@ step.
Numbers of constraints and variables for this ILP are presented
	adjacent to the same statistics for the (single) ILP in the existing approach.
For the smaller, simpler architectures (\adres@ \& \clustered@), the ILP sizes are similar.
We believe that the reason solving this ILP is faster than
	the existing approach is
	the inherent flexibility of allowing shorts:
	an initial guess for any variable is more likely to result in a solution.
We also note that this work scales better:
	while the constraint numbers for the
	existing approach follow the trend of the MRRG size,
	this approach remains more constant, as its more dependent on number of FUs.
Models for mapping to \hycube@ are significantly smaller than \adres@ and Clustered,
	due to it requiring a lower \numnei@, which results in
	many fewer path variables and associated constraints.
Statistics for the placement-only and routing-only ILP of this work are also included underneath,
	and are one to two orders-of-magnitude less than the \relaxedPathConstraintPlacementName@ ILP.


\section{Related Work}
There have been a few proposed integer linear programs for CGRA mapping
	\cite{CGRAME-ILP,CGRA-ILP-QEA,GeneralConstraintSpatial}, and a SAT-based one \cite{WaveSAT}.
\cgrame@'s existing approach \cite{CGRAME-ILP} as well as \cite{GeneralConstraintSpatial} are the most general,
	deferring all scheduling and mapping decisions to the ILP/SAT solver,
	and modelling routing explicitly.
All these approaches attempt to completely retain the flexibility of the
	architectures that they support,
	but also suffer from long runtimes.
In the case of \cite{WaveSAT},
	it was necessary to break up the SAT problem in time,
	so that the runtime remained bounded,
	via an interesting sliding-window approach.
Our approach
	builds upon some ideas from these works,
	applying the new idea of a systematically simplified device model.

While the ILP and SAT methods mentioned above try to solve placement
	and routing at the same time, this work splits them up somewhat,
	like \cite{SPR,DRESC,ScalableMapCGRA,CGRA-ILP-QEA}
	or typical FPGA mapping approaches.
The aim is to break-up one intractable problem into smaller tractable problems,
	but CGRAs have resisted this by having inflexible routing, as evidenced by the long runtimes of \cite{SPR,DRESC}.
The approach presented here is more of a hybrid,
	explicitly modelling hard routing constraints during placement,
	and then checking feasibility in a detailed routing step.

This approach is influenced by \cite{EPIMap,REGIMap,EcMS,SCCbMS,SchedBindRouteOpBased},
	in that it primarily operates on connectivity information,
	but these other mapping procedures are each tuned/designed for a specific class of architectures.
The methods of \cite{REGIMap, EPIMap, CGRA-ILP-QEA,ScalableMapCGRA}
	manipulate the DFG to assist in mapping the architectures
	(such as node duplication, and adding explicit ``routing'' nodes).
To remain generic, we hesitate to manipulate the DFG,
	instead allowing scheduling to be decided as part of the placement.

\section{Conclusion \& Future Directions}
We have presented an ILP-based method for mapping applications to a variety of CGRAs
	in a reasonable amount of time, with significant improvement over the state-of-the-art in generic CGRA mapping \cite{CGRAME-ILP}.
To extend this work, the derived connectivity information
	could instead be applied to a polynomial-time approach generalized from one of \cite{REGIMap, EPIMap, SCCbMS}.
A separate direction is to not solve the entire problem at once:
	partition the DFG and/or CGRA, or use the ``sliding window'' approach of \cite{WaveSAT}.
However, partitioning and incremental techniques are difficult to do generically,
	and must be done with care to take advantage of the capabilities of the architecture.
As discussed in \cref{sec:exp:discussion},
	reducing the number of path variables is essential to the performance of this work,
	so heuristic methods for achieving this are of interest.

\section*{Acknowledgement}
\ifCLASSOPTIONpeerreview(empty for peer review)\else
The authors gratefully acknowledge Huawei for supporting this research.
Computations were performed on the Niagara supercomputer at the SciNet HPC Consortium. \cite{scinet-citeme}
\fi

\printbibliography
\ifCLASSOPTIONpeerreview\else\fi
\ifIEEEXplore
	\end{NoHyper}
\fi
\end{document}